\documentclass[default,iicol]{sn-jnl}

\usepackage{graphicx}%
\usepackage{multirow}%
\usepackage{amsmath,amssymb,amsfonts}%
\usepackage{amsthm}%
\usepackage{mathrsfs}%
\usepackage[title]{appendix}%
\usepackage{xcolor}%
\usepackage{textcomp}%
\usepackage{manyfoot}%
\usepackage{booktabs}%
\usepackage{algorithm}%
\usepackage{algorithmicx}%
\usepackage{algpseudocode}%
\usepackage{listings}%

\usepackage{scrtime}
\usepackage[bottom]{footmisc}

\usepackage{hyperref}
\hypersetup{
colorlinks=true, 
citecolor=magenta,
breaklinks=true, 
urlcolor= blue, 
linkcolor= blue, 
bookmarksopen=true, 
}

\usepackage{natbib}
\usepackage{color}
\usepackage[type={CC},modifier={by-nc-sa},version={4.0},]{doclicense}
\definecolor{linkcolor}{rgb}{0,0,0.6}

\newcommand{\revise}[1]{{\leavevmode\color{black}#1}}
\newcommand{\reviseCTP}[1]{{\leavevmode\color{black}#1}}

\newcommand{\Rset}{\mathbb{R}}
\newcommand{\ed}{\mathrm{e}}

\newcommand{\id}{\mathrm{i}}
\newcommand{\boldsigma}{\boldsymbol{\sigma}}

\DeclareMathOperator{\am}{am}
\DeclareMathOperator{\sn}{sn}
\DeclareMathOperator{\ellipticK}{K}

\DeclareMathOperator{\Tr}{\textup{tr}}

\begin{document}

\title[A numerical direct scattering  method for the periodic sine-Gordon equation]{A numerical direct scattering method for the periodic sine-Gordon equation}


\author[1]{\fnm{Filip} \sur{Novkoski}}\email{filip.novkoski@u-paris.fr}
\equalcont{These authors contributed equally to this work.}

\author[1]{\fnm{Eric} \sur{Falcon}}\email{eric.falcon@u-paris.fr}
\equalcont{These authors contributed equally to this work.}


\author*[3]{\fnm{Chi-Tuong} \sur{Pham}}\email{chi-tuong.pham@upsaclay.fr}
\equalcont{These authors contributed equally to this work.}

\affil[1]{Université Paris Cité, CNRS, MSC, UMR 7057, F-75013 Paris, France}
\affil[3]{Université Paris-Saclay, CNRS, LISN, UMR 9015, F-91405 Orsay, France}



\abstract{ We propose a procedure for computing the direct scattering
  transform {of} the periodic sine-Gordon equation. This procedure,
  previously used within the periodic Korteweg--de Vries equation
  framework, is implemented for the case of the sine-Gordon equation
  and is validated numerically. In particular, we show that this
  algorithm works well with signals involving topological solitons,
  such as kink or anti-kink solitons, but also for non-topological
  solitons, such as breathers. It has also the ability to distinguish
  between these different solutions of the sine-Gordon equation within
  the complex plane of the eigenvalue spectrum of the scattering
  problem. The complex trace of the scattering matrix is made
  numerically accessible, and the influence of breathers on the latter
  is highlighted. Finally, periodic solutions of the sine-Gordon
  equation and their spectral signatures are explored in both the
  large-amplitude (cnoidal-like waves) and low-amplitude (radiative
  modes) limits.}




\maketitle

\insert\footins{
  \normalfont\footnotesize
  \interlinepenalty\interfootnotelinepenalty
  \splittopskip\footnotesep \splitmaxdepth \dp\strutbox
  \floatingpenalty10000 \hsize\columnwidth
   \doclicenseText  \doclicenseImage[imagewidth=4em]\par}


\section{Introduction}

Soliton are encountered in various fields of physics, from condensed matter, optics to
hydrodynamics~\cite{Russell1844,Frenkel1939,RemoissenetBook,DauxoisBook}. It has also received
considerable attention in applied mathematics, in particular, in the study of integrable
systems~\cite{Ablowitz1981,Drazin1989}. Advanced tools of the inverse scattering transform (IST)
have significantly changed the theoretical view of soliton solutions of nonlinear partial
differential equations (PDEs), such as Korteweg--de Vries (KdV) and nonlinear Schr\"odinger (NLS)
equations~\cite{OsborneBook}. It has also recently led to experimental applications in
optics~\cite{Suret2023Refraction} and in hydrodynamics surface waves~\cite{Redor2019}, such as the experimental synthesis of
soliton gas~\cite{Suret2020} or Peregrine solitons~\cite{TikanPRF22a}.

The sine-Gordon (SG) equation admits a variety of different soliton solutions, such as topological
solitons (in the form of kinks or anti-kinks), non-topological breathers (i.e., solutions
oscillating in time and localized in space), or multisolitons. Sine-Gordon solitons have been
experimentally evidenced in the canonical example of a chain of coupled pendulums~\cite{Scott1969},
but also in Josephson tunnel junctions between superconductors~\cite{Ustinov1998}, dislocations in
crystals~\cite{Frenkel1939,DauxoisBook}, or in biological cellular structures~\cite{Ivancevic2013}.

The SG equation has been extensively studied in the context of the IST~\cite{Malomed2014},
and, in particular, \revise{the initial value problem can be solved using IST
methods}~\cite{Ablowitz1973,Drazin1989}. The corresponding Lax pairs are well
known~\cite{Takhtadzhyan1974}, whereas the spectral theory behind the SG equation with
periodic boundary conditions has been developed using Floquet theory~\cite{Forest1982}.

Generally speaking, if an integrable equation admits soliton solutions, a specific complex value
$\lambda$ can be assigned to each of such solitons, thus fully characterizing the solution. The
value $\lambda$ is the solution of a linear eigenvalue problem related to the given integrable
nonlinear PDE. For example, in the case of the KdV equation, the linear problem is related to the
Schr\"odinger equation, $\psi_{xx}-(u-\lambda)\psi=0$, of the field $\psi(x,t)$ with an unknown
potential $u(x,t)$ coming from the solution of the KdV solution~\cite{Gardner1967}. \revise{A
  critical property of integrable equations is isospectrality, meaning that the values of $\lambda$
  are constant in time.} The direct oscattering problem aims to seek the values of $\lambda$
(independent of time) from a given initial condition $u(x,0)$. For the KdV equation, the values of
$\lambda$ are purely real, whereas, for the \revise{focusing} NLS and SG equations, $\lambda$ can take also complex
values as well.\revise{While on the infinite line, this eigenvalue is unique, as we will see, in the
  periodic case this corresponds to a band which is bounded by two values.} Numerical simulations of
the eigenvalue equations of the direct scattering in the sine-Gordon framework have been performed,
as well as the IST by numerically solving the related nonlinear eigenvalue problem to achieve the
soliton spectrum ~\cite{Overman1986,Flesch1991}. Here, we apply a more general and straightforward method for
the direct scattering problem since it comes down to a product of matrices. Such a method has
been previously applied to the KdV equation~\cite{OsborneBook}{, and used experimentally in the
  periodic KdV equation~\cite{Novkoski2022}}.


While SG solitons have been experimentally observed in various
fields~\cite{Scott1969,Ustinov1998,Ivancevic2013}, a direct link between the direct scattering
method and experimental data is still elusive so far. An experimental investigation of the
integrability of physical systems described by the SG equation is yet to be made and would also
provide a step towards the experimental study of integrable turbulence, in particular in the case of
topological solitons, and for the synthesis of corresponding soliton gas. Here, we present a tool
and a case study of individual solutions to the SG equation, that could potentially allow for easy
use of the direct scattering method on experimental data.

The article is organized as follows: in Sec.~\ref{realSGE}, we compare classic solutions of the SG
equation on the real line to those satisfying periodic boundary conditions. In Section
\ref{DSmethod}, we recall the scattering theory generally used in the literature and introduce a
simpler numerical scattering method. We then apply the latter to well-known solutions and test its
validity (Sec. \ref{sol} for one- and two-kink solitons, along with breathers ; Sec.
\ref{periodic_sol} for periodic trains of kinks and radiative modes).  We draw our conclusions in
Sec.~\ref{conclusion}.

\section{The sine-Gordon equation on the real line and with periodic boundary conditions} \label{realSGE}
\subsection{On the real line} The (1+1) sine-Gordon equation \revise{(1 dimension of space and 1 of time)} on
the real line as an initial-value problem and for the unknown field $\phi(x,t)$, reads in
nondimensional form
\begin{equation}\label{eq:sg}
  \phi_{tt}-\phi_{xx}+\sin\phi=0,
\end{equation}
for $(x,t) \in \Rset\times[0,\infty]$ and an initial condition $\phi(x,t=0) = \phi_0(x)$.

One set of spatial boundary conditions can be chosen as ``vanishing'' ones at infinity, that is
\begin{align}
  &\phi(x,t) \to 0 \pmod{2\pi}, \quad \text{as}\; |x|\to \infty \label{vanishingBC1} \\
  &\phi_x(x,t) \to 0, \quad \text{as}\; |x|\to \infty. \label{vanishingBC2}
\end{align}
In this article, we will refer to this SG equation along the real line, with vanishing boundary
conditions as the iSG equation.

A well-known solitonic solution of this problem reads
\begin{equation}
  \phi^{\sigma}_{\mathrm{kink}}(x,t) = 4\arctan \exp\left[ \sigma \frac{x-vt-x_0}{\sqrt{1-v^2}}\right]. \label{eq:formule_kink}
\end{equation}
The case $\sigma = +1$ corresponds to the so-called kink solution, whereas the case $\sigma = -1$ corresponds
to the so-called anti-kink solution, both moving at speed $v\in ]-1, +1[$.

\reviseCTP{Note that throughout the rest of the article, we will refer to the term kink (resp. anti-kink) as a topological state  connecting values of the function $\phi(x,t)$ going from $2n\pi$ to $2(n+1)\pi$ (resp. $2m\pi$ to $2(m-1)\pi$) as $x$ increases, with $m, n$ integers.} 

If the vanishing boundary conditions (\ref{vanishingBC1}, \ref{vanishingBC2}) are disregarded,
another set of solutions, moving at speed $v \in\, ]-1, +1[$, can be found as an infinite periodic
train of kinks ($\sigma = +1$) or anti-kinks ($\sigma = -1$). They read
\begin{equation}
  \phi^{\sigma}_{\am}(x,t) = \pi + \sigma \am\left(\frac{x-vt-x_0}{\sqrt{m(1-v^2)}}, m\right), \label{formule_am} 
\end{equation}
\revise{where $\mathrm{am}(x,m)$ is the Jacobi amplitude function,} for $m \in\, ]0, +1[$. They form a
two-parameter family of solutions, parameterized by their speed $v$ and the elliptic parameter
$m$. $\sigma = +1$ is related to the increasing solution (kinks), whereas $\sigma = -1$ corresponds
to the decreasing one (anti-kinks). Note that the limit-case $m=1$ corresponds to the kink or the
anti-kink solutions given by \eqref{eq:formule_kink}.

For such a solution, the \revise{period} $\Lambda$ between two neighboring periodic kinks is given by
\begin{equation}
  \Lambda(m,v) = 2 \sqrt{m(1-v^2)} \ellipticK(m), \label{distance_interkink}
\end{equation}
{where $\ellipticK$ is the complete \revise{elliptic} function of the first kind. }

\subsection{Periodic boundary conditions} 
\begin{figure}[t!]
  \centering

  \includegraphics[width=0.85\columnwidth]{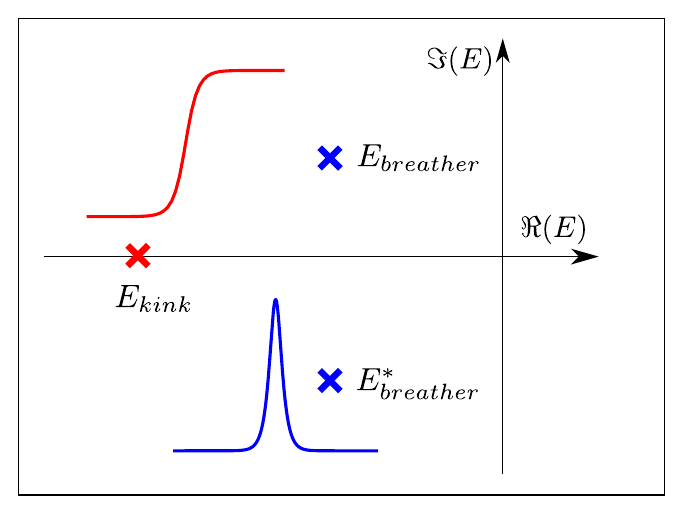}
  \caption{Scheme in the complex plane of the energies, $E=\lambda^2$, of different
    types of soliton that can take real (topological soliton in red) or complex (breather soliton in blue) values.}\label{fig:eigenvalues}
\end{figure}

In the rest of the article, we study the (1+1)
SG equation \eqref{eq:sg} 
but we impose ``periodic'' boundary conditions with a spatial period $L$ :
\begin{align}
    &\phi(x+L,t) = \phi(x,t)  \pmod{2\pi}, \label{periodicBC1}\\
    &\phi_x(x+L,t) = \phi_x(x,t), \label{periodicBC2}
\end{align}
We assume the initial condition $\phi(x,t=0)=u(x)$, as well as $\phi_x(x,t=0)$ and $\phi_t(t=0)$ are
known. We will call this set of equations the periodic sine-Gordon equation (pSG). The classic
sine-Gordon equation is known as the continuous limit of a discrete mechanical chain of torsionnally
coupled pendula. Here, the periodic conditions (\ref{periodicBC1}, \ref{periodicBC2}) turns the
usual straight pendulum chain into a circle chain (see Top of Fig. \ref{fig:eigenvalues}) and
includes possible non-zero topological charges, hence the possible presence of kinks.

The length $L$ will be chosen to be larger than the typical length scale equal to $1$ in the case of
the nondimensional SG equation \eqref{eq:sg}.

The equivalent of the 1-kink solution for such periodic boundary conditions and moving at speed $v$
corresponds to the solution given by \eqref{formule_am}, provided the distance between two successive periodic kinks is
$L$, namely we have the relationship between $L, v, m$
\begin{equation}
  L = \Lambda(m,v) = 2 \sqrt{m(1-v^2)} \ellipticK(m). \label{eq:Lmv}
\end{equation}
In turn, the $N$-periodic kink solutions \reviseCTP{(in the sense of a $2N\pi$-jump in $\phi$)} for such periodic boundary conditions are given by
\eqref{formule_am}, with the relationship between $L, v, m, N$
\begin{equation}
  L = N \Lambda(m,v) = 2 N \sqrt{m(1-v^2)} \ellipticK(m). \label{eq:LmvN}
\end{equation}

\section{Direct scattering method} \label{DSmethod}

From the solutions of the pSG equation, we wish to obtain the nonlinear spectral content of our system.
{To that end, the SG equation is viewed as a} compatibility condition of two matrix equations~\cite{Malomed2014}
\begin{equation}
  \begin{aligned}
  \Psi_x&=\id\left[\frac{\lambda}{2}\boldsigma_z+\frac{\phi_x-\phi_t}{4}\boldsigma_x-\frac{\boldsigma_z\cos\phi+\boldsigma_y\sin\phi}{8\lambda}\right]\Psi \label{eq:compatibility}, \\
    \id\Psi_t&=\left[\frac{\lambda}{2}\boldsigma_z+\frac{\phi_x-\phi_t}{4}\boldsigma_x+\frac{\boldsigma_z\cos\phi+\boldsigma_y\sin\phi}{8\lambda}\right]\Psi,
  \end{aligned}
\end{equation}
where $\Psi(x,t)$ is a two-component complex Jost function, and $\boldsigma_j$ are the Pauli
matrices with $j=x$, $y$ or $z$
  \begin{equation}
    \boldsigma_x =  \begin{bmatrix} 0 & +1 \\ -1 &0 \end{bmatrix}; \boldsigma_y =  \begin{bmatrix} 0 & -\id \\ +\id &0 \end{bmatrix}; \boldsigma_z =  \begin{bmatrix} +1 & 0 \\ 0 & -1  \end{bmatrix}.
  \end{equation}
  
  Unlike the KdV or NLS equation, where the problem of finding $\lambda$ is essentially linear, here
  {we are dealing with} an \revise{eigenvalue problem which is linear in $\Psi$ but nonlinear in
    $\lambda$}. We will be interested in finding the value $E$, which we refer to as
  \textit{energy}, such that $\lambda=\sqrt{E}$.

{Because of the periodic boundary conditions,} 
it is possible to gain more insight of the problem by applying Floquet 
theory~\cite{Forest1982}. {After fixing a given point $x_0$, we consider} a basis of solutions
$\{\Psi_+(x,x_0,E),\Psi_-(x,x_0,E)\}$ \revise{such that
  \begin{align}
    \Psi_+(x=x_0,x_0,E)&=
    \begin{bmatrix}
      1 \\ 0
    \end{bmatrix}, \\
    \Psi_-(x=x_0,x_0,E)&=\begin{bmatrix}
      0 \\ 1
    \end{bmatrix}.
  \end{align}
}
Due to the periodicity, the functions $\Psi_+(x+L,x_0,E)$ also
have to be solutions of Eq.~\eqref{eq:compatibility}, and are expressible in terms of the basis
functions. This leads us to write
\begin{align}
  \begin{bmatrix}
    \Psi_+(x+L,x_0,E)\\ \Psi_-(x+L,x_0,E)
  \end{bmatrix}
  = \mathbf{S}(E)\begin{bmatrix}
    \Psi_+(x,x_0,E)\\ \Psi_-(x,x_0,E)
  \end{bmatrix}
\end{align}
where $\mathbf{S}$ is the scattering matrix ({or transfer matrix, }also called the {\em monodromy matrix}). In order for the
solutions to be bounded, certain conditions need to be imposed onto the matrix. In addition, the
spectrum of the problem in Eq.~\eqref{eq:compatibility} is divided into its \textit{discrete} and
\textit{continuous} parts.

\revise{The eigenvalues of the discrete spectrum correspond to solitons and are such that they satisfy~\cite{Forest1982}
\begin{gather}
\left|\Tr (\mathbf{S})\right|=2, \\ 
\Im(\Tr \mathbf{S})=0, 
\end{gather}}
where $\Im$ stands for the imaginary part. In addition,
for real potentials, the eigenvalues are negative. Thus, the solitons will lie in the $E<0$ plane as
shown in Fig.~\ref{fig:eigenvalues}, for the case of kinks (for which $E\in\mathbb{R}$) and
breathers ($E\in\mathbb{C}\setminus \mathbb{R}$).

{On the other hand, }the continuous spectrum consists of eigenvalues such that $E>0$ with
\revise{
\begin{gather}
  \left|\Tr (\mathbf{S})\right|\leq 2,\\
  \Im(\Tr \mathbf{S})=0.
\end{gather}}
This part of the spectrum is associated with the radiation of phonons,
i.e., the dispersive component of the wave field. Thus, in order to spectrally detect the solitons (in the
$E$-plane), { numerical computations of the trace of the monodromy matrix \revise{are} needed}.

{In this article, we tackle the problem by following a different approach, using }a method similar to that first used in
Refs.~\cite{Osborne1994,OsborneBook} for the periodic KdV case. Expanding the Jost function at
$x+\Delta x$ yields
\begin{equation}\label{eq:exponent}
  \Psi(x+\Delta x)=\ed^{\mathbf{T}\Delta x}\Psi(x)=\mathbf{Q}(x)\Psi(x),
\end{equation}
where the matrix $\mathbf{T}$ is inferred from Eq.~\eqref{eq:compatibility} and $\mathbf{Q}$ is
defined as $\mathbf{Q}(x)\equiv \ed^{\mathbf{T}\Delta x}$. By iterating the above process over a
piecewise constant signal containing $N$ points, we find the scattering matrix
\begin{equation}\label{eq:product}
  \mathbf{M}(E)=\prod_{n=N-1}^0 \mathbf{Q}(x_n,E),
\end{equation}
which relates now $\Psi(x+L)$ to $\Psi(x)$, \revise{with $x_n=n\Delta x$ and $\Delta x=L/N$}. By choosing
$x=x_0$, we have that $\mathrm{tr}(\mathbf{M})=\mathrm{tr}(\mathbf{S})$.
\reviseCTP{Note that our calculation of $\mathbf{M}(E)$ consists in a mere product of matrices, whereas the computations of $\mathbf{S}(E)$ is based on non-straightforward numerical schemes~\cite{Overman1986}.}

In order to simplify computations, we are able to reduce the matrix $\mathbf{Q}$ of
Eq.~\eqref{eq:exponent} to
\begin{align}
  \mathbf{Q}&=\cos(\Delta x)\mathbb{I}+\id\sin(\Delta x)\vec{\mathbf{v}}\cdot\vec{\pmb{\sigma}}, \\
  \mathrm{with\quad}\,\vec{\mathbf{v}} &= \Big(\frac{\phi_x-\phi_t}{4},-\frac{\sin\phi}{8\lambda},\frac{\lambda}{2}-\frac{\cos\phi}{8\lambda}\Big),
\end{align}
speeding calculations up around two times. \revise{In practice, in order to compute $\mathbf{Q}$, we
  assume that $\phi(x,t=t_0)$ as well as $\phi(x,t=t_0+\Delta t)$ are known, since one only has to have
  information of the time derivative at $t_0$. Knowing the
  full temporal evolution of $\phi$ is unnecessary.} Despite this simplification, computation using complex
values is still required. A decomposition of Eq.~\eqref{eq:product} into real and imaginary parts
would be a further improvement, but is not obvious.

{In the following, using well-known solutions along with their spectral characteristics, we will test this numerical method and compare its results to those of the literature (for instance those studied in Ref.~\cite{Forest1982} using the $\mathbf{S}$ matrix).
}

\section{One- and two-soliton solutions}\label{sol}
\subsection{\reviseCTP{Single} kink solitons for periodic boundary conditions}
\begin{figure}[t!]
  \centering
  \includegraphics[width=\columnwidth]{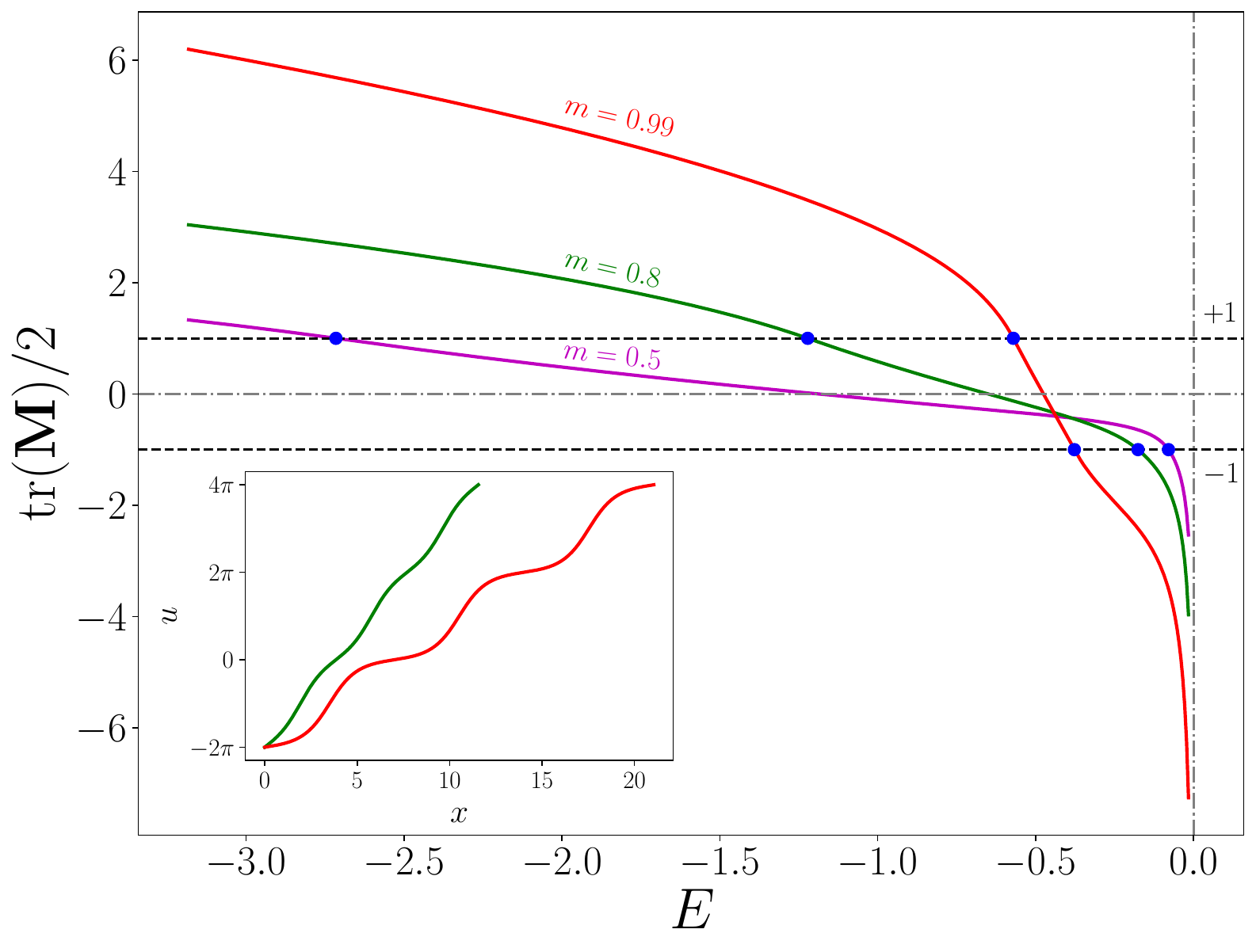}
  \caption{\label{fig:trace-periodic}Half-trace of the monodromy matrix $\mathbf{M}(E)$ for
    three kink solitons with p.b.c, $m=0.5,0.8,0.99$. \revise{The trace crosses the lines $\pm 1$ only once
    respectively for all three.} We observe how the energies move closer together as $m$ moves
    towards $1$. \reviseCTP{Inset: Single kink waveform $\phi(x,0)=u(x)$ given by Eq.~\eqref{formule_am} and used as the initial condition for
    the direct scattering method with $m=0.8$ and $m=0.99$ (same color as in the main figure). Note that,
    in both cases, we have used three periods $L(m)$ (see Eq.~\eqref{eq:Lmv}) as an illustration of their periodicity (hence a $6\pi$-jump).}}
\end{figure}

{Let us begin with the topological soliton solutions of the pSG equation, i.e, periodic kinks and
anti-kinks (see Eqs. \eqref{formule_am} and \eqref{eq:Lmv}). We examine three solitons, with the
same velocity $v=0.3$, but different values of the elliptic parameter $m=0.5,0.8,0.99$, described by
Eq.~\eqref{formule_am}. Using the algorithm prescribed above, {based on the calculation of the matrix $\mathbf{M}(E)$, we use these three soliton solutions} to compute
their corresponding traces of the scattering matrix. We focus on kinks, i.e. $\sigma=+1$, and the
space domain is chosen so that it corresponds to one period leading to one single soliton. {As a consequence,  each spatial domain has a different length $L$}.

The results for the traces are shown in Fig.~\ref{fig:trace-periodic}. We observe that for all those three
solitons, there are one intersection with $+1$ and one with $-1$ as expected theoretically. These
intersections corresponding to solitons, we see that the method permits their detection. As the
elliptic parameter is varied, the two corresponding energies move closer and closer together (and in
the limit of $m=1$, they would coincide). The velocity $v$ and elliptic parameter $m$ are linked to
the energies of the  soliton and are given by
\begin{align}\label{eq:train_velocity}
  m&=\frac{2}{1+\tfrac{1}{2}\left[\sqrt{{E_2}/{E_1}}+\sqrt{{E_1}/{E_2}}\right]},\\
  v&=\frac{4(E_1E_2)^{1/2}-1}{4(E_1E_2)^{1/2}+1},
\end{align}
\revise{as computed using expressions found in~\cite{Forest1982}}. It is worth mentioning that previous
references dealing with this type of solutions do not state that the elliptic parameter $m$ is fixed
by the energies in this way. Using a bisection method, we numerically determine the intersections
with $\Tr (\mathbf{M})=\pm 1$, allowing us to obtain the energies \revise{$E_1$ and $E_2$}. {Given
  this statement, along with Eq.~\eqref{eq:train_velocity},} we are able to confirm that the method
correctly identifies the solitons, with the numerical error being within $0.3\%$. }

\begin{figure}[t!]
  \centering
  \includegraphics[width=\columnwidth]{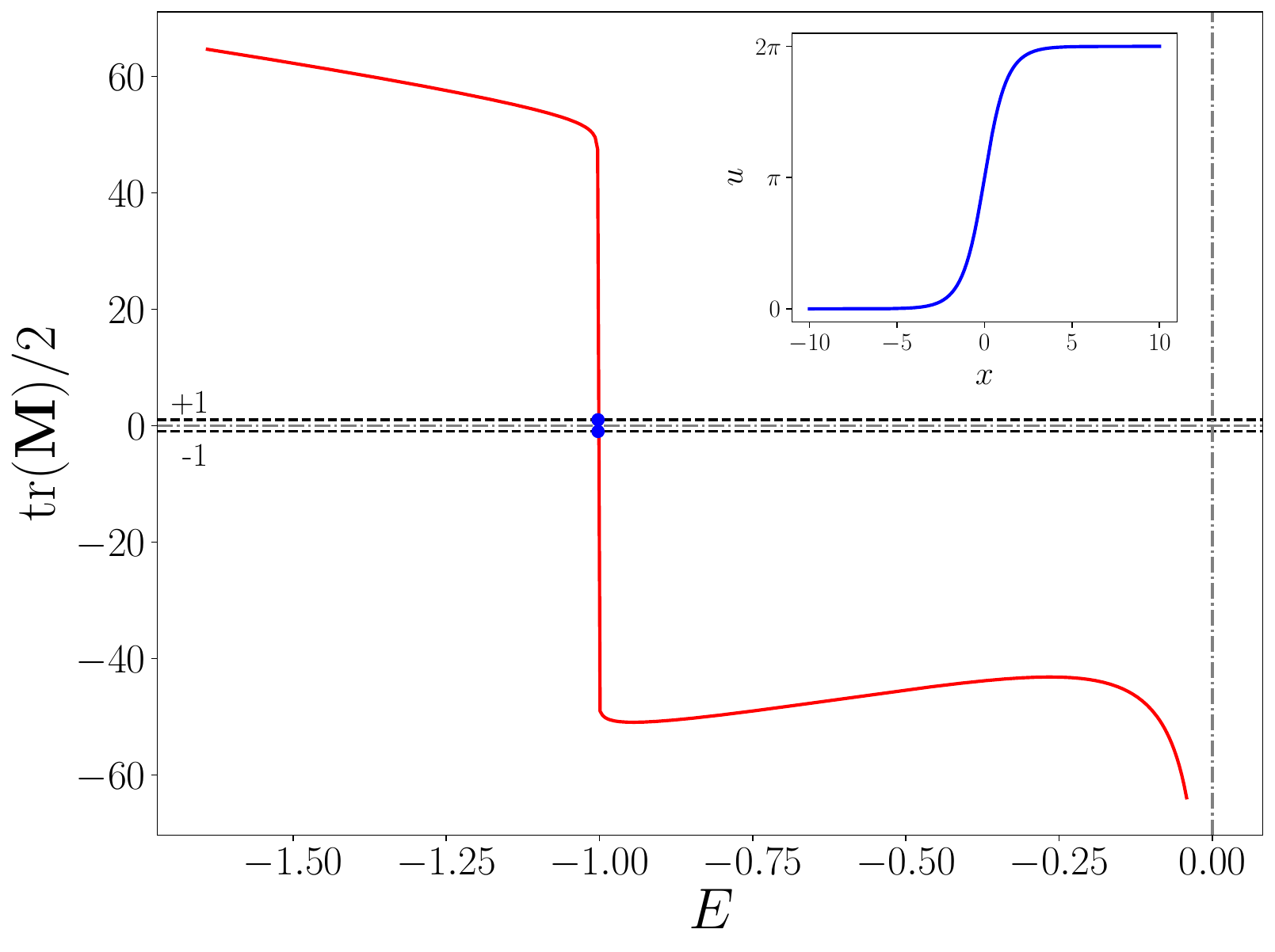}
  \caption{\label{fig:trace-kink}Half-trace of the monodromy matrix $\mathbf{M}(E)$ for a kink
    soliton. Only one crossing of the trace at $\pm1$ is found for a negative value of the energy,
    $E=-1$, as expected theoretically. Inset: The kink-soliton waveform $\phi(x,0)=u(x)$ used as the
    initial condition for the direct scattering method (Eq.~\eqref{eq:formule_kink}, with $L=20$).  }
\end{figure}

\subsection{Truncated kink soliton of the whole line}
Here, we will truncate the kink solution on the real line.  We want to show in what extent the
numerical procedure still holds when the input signal is not an exact solution (here, the boundary
conditions (equations \eqref{periodicBC1} and \eqref{periodicBC2}) are not satisfied). Let us begin with the
topological soliton solutions of the iSG equation, i.e, kinks and anti-kinks.

A single soliton solution with energy $E=-K$ is given by  \eqref{eq:formule_kink}
and its velocity $v$ satisfies
\begin{equation}\label{kinkvelo}
  v=\frac{4K-\sigma}{1+4K\sigma},
\end{equation}
for $\sigma=\pm 1$ discriminating kinks ($+$) from anti-kinks ($-$)~\cite{Kivshar1991} .

We apply the algorithm proposed in Sec.~\ref{DSmethod} in order to compute the trace of the
scattering matrix $\mathbf{M}$ and to find the eigenvalues $\lambda$ of the following signal. The
input signal is chosen to be a kink (displayed in the inset of Fig.~\ref{fig:trace-kink}), for which
we take $K=1$ in Eq.~\eqref{kinkvelo}, $\sigma=+1$, and $u(x)=\phi(x,0)$ as in Eq.~\eqref{eq:formule_kink}
consisting of $N=1000$ points, on an interval {of length $L=20$. While the signal} is a solution for the
infinite line, and does not satisfy perfectly the periodic boundary condition, we consider that the
errors due to the mismatch at the two ends will be negligible and use the given form of the field
nevertheless. By iterating over both the signal and different values for the energy $E$, we find the
half-trace of the matrix $\mathbf{M}(E)$ as shown in Fig.~\ref{fig:trace-kink}. The imaginary part
of the trace is zero, and the real part of $\mathrm{tr}(\mathbf{M})/2$ takes values $\pm 1$ at
$E_1=E_2=-1.0024$, as expected. These values of the crossings are obtained using the bisection
method with the cutoff precision set to $10^{-12}$. We do not notice a numerical difference between
the two values.  Close to $E=0$ the trace tends to $-\infty$ which is expected theoretically
\cite{Forest1982}. {In conclusion, the algorithm has detected the presence of a kink soliton on a line, without
imposition of periodic boundary conditions.}

\subsection{\revise{Two-kink solution}} We apply now the procedure for a signal consisting of a pair of two kinks with two different energies $K_1$ and $K_2$ and
their corresponding velocities $v_1$ and $v_2$. The input signal is calculated using the formula for a
two-kink solution that can be obtained through a B\"acklund transform~\cite{Barone1971} and reads
\begin{gather}
  u(x,t)=4\arctan\left(k\frac{\phi_1-\phi_2}{1+\frac{\phi_1}{\phi_2}}\right),\label{Backlund}\\
  \phi_j=\exp\left(\frac{x-v_jt}{\sqrt{1-v_j^2}}\right), \quad v_j=\frac{4K_j-1}{1+4K_j},\\
  k=\frac{\beta_1-\beta_2}{\beta_1+\beta_2}, \quad \beta_j=\frac{1-v_j}{1+v_j}, \quad j=1,2 .
\end{gather}
We here stress that the link between the eigenvalues, i.e. $K_1$ and $K_2$, and the
velocities ($v_1$ and $v_2$) is the same as for the one-kink solution. This link has not been clearly stated in the
literature, to our knowledge.

\begin{figure}[t!]
  \centering
  \includegraphics[width=\columnwidth]{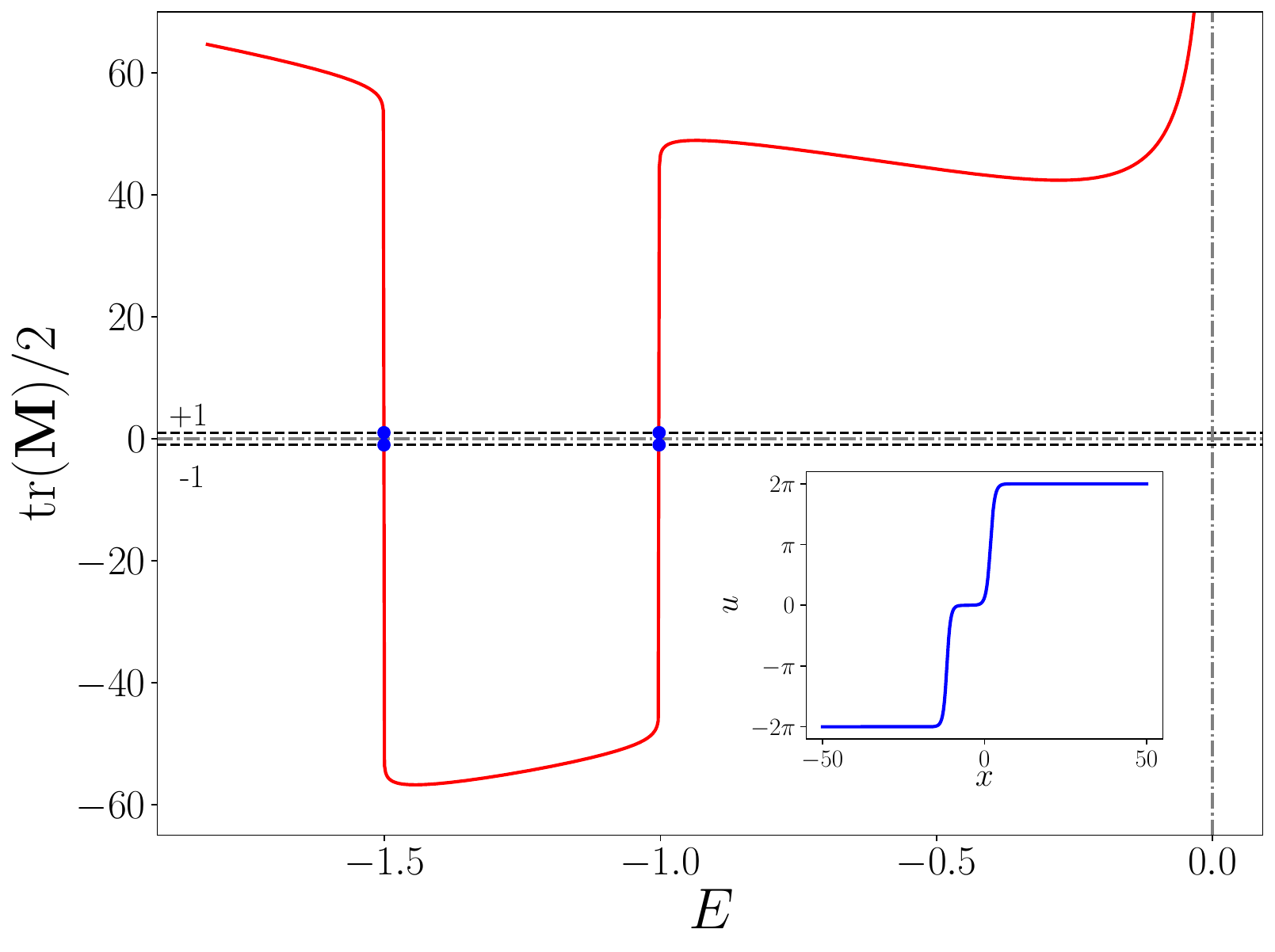}
  \caption{\label{fig:trace-kink-pair}Half-trace $\mathbf{M}(E)$ for a pair of kink solitons with different
    energies. As predicted theoretically, two crossings at the negative energies are found,
    corresponding well numerically to those prescribed. $L=100$. Inset: Signal used as the initial condition $u(x)$ in the direct
  scattering, with energies $E=-1$ and $E=-1.5$ (see Eq.~\eqref{Backlund}).}
\end{figure}

The initial condition used is shown in the inset of Fig.~\ref{fig:trace-kink-pair}) for $L=100$,
with the energies of the two solitons fixed {at values} $K_1=1$ and $K_2=1.5$. In response to this initial
condition, the numerically computed half-trace $\mathbf{M}(E)$ is shown in
Fig.~\ref{fig:trace-kink-pair}. The energies found using the bisection method lead to
$E_1=E_2=-1.5009$ and $E_3=E_4=-1.0024$, which is in {excellent} agreement with the prescribed initial soliton
energies. The application of the algorithm to arbitrary numbers of kink or anti-kink solitons is
thus achievable as long as their energies are different enough.  Due to the fact that their energies are real numbers, the detection of kinks and
anti-kinks is numerically easy, and the computation time is relatively short.
{The case of a kink train will be
tested below in section \ref{periodic_sol}}.

\begin{figure*}[t]
  \centering
  \quad(a) \hspace*{7.5cm}(b)
  
  \includegraphics[width=\columnwidth]{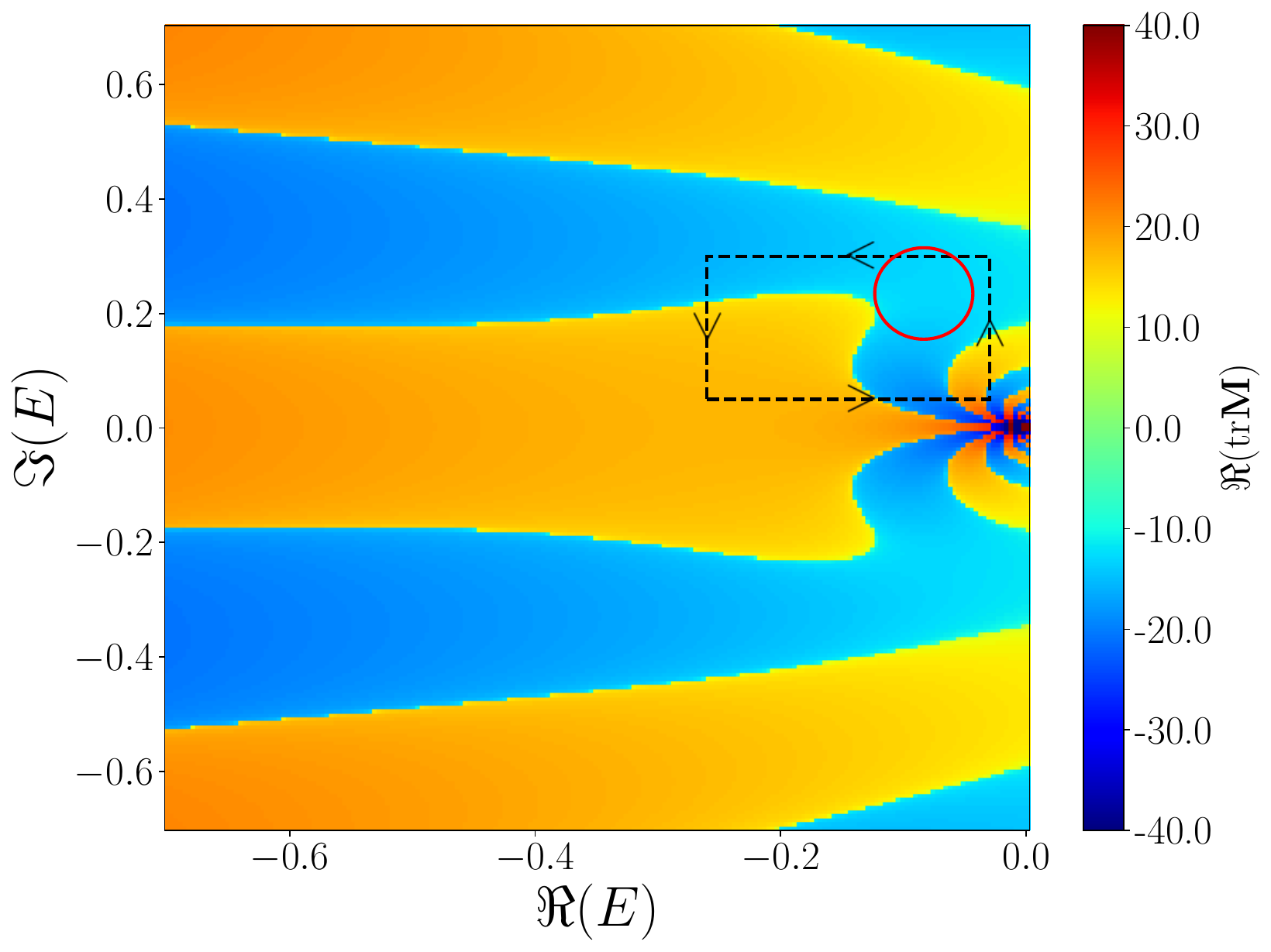}
  \includegraphics[width=\columnwidth]{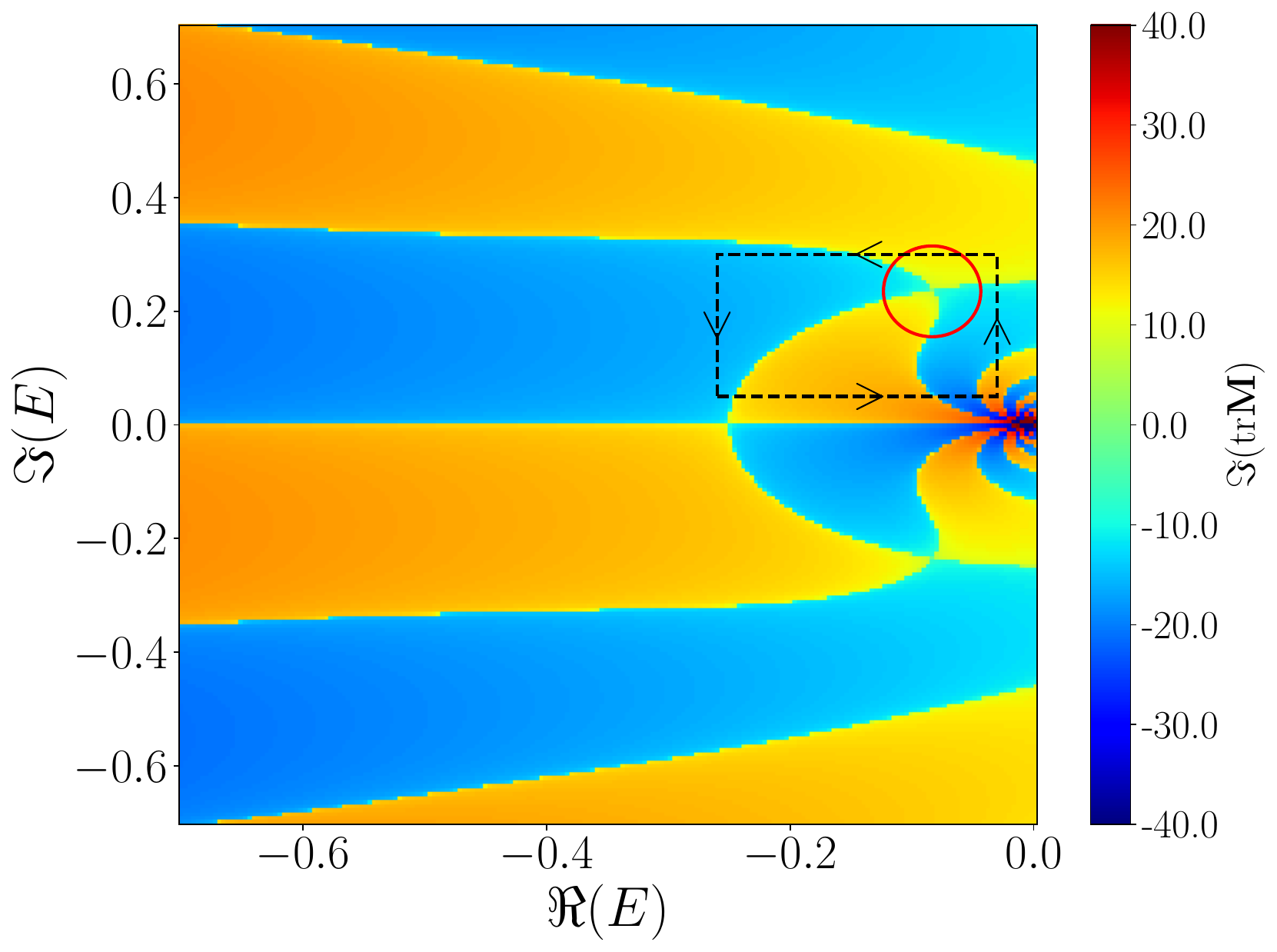}
  \caption{\label{fig:trace-breather}(a) Real part of the trace of $\mathbf{M}(E)$ showing
    only a slight deformation close to the complex value of the breather energy.
    (b) Imaginary part of the trace of $\mathbf{M}(E)$ in the
    complex plane of energy $E$ demonstrating a strong ``pinch'' close to the breather energy
    $E=\frac{1}{4}\ed^{2\id\pi/3}$ (red circle).}
\end{figure*}

\begin{figure}[t!]
  \centering
  \includegraphics[width=\columnwidth]{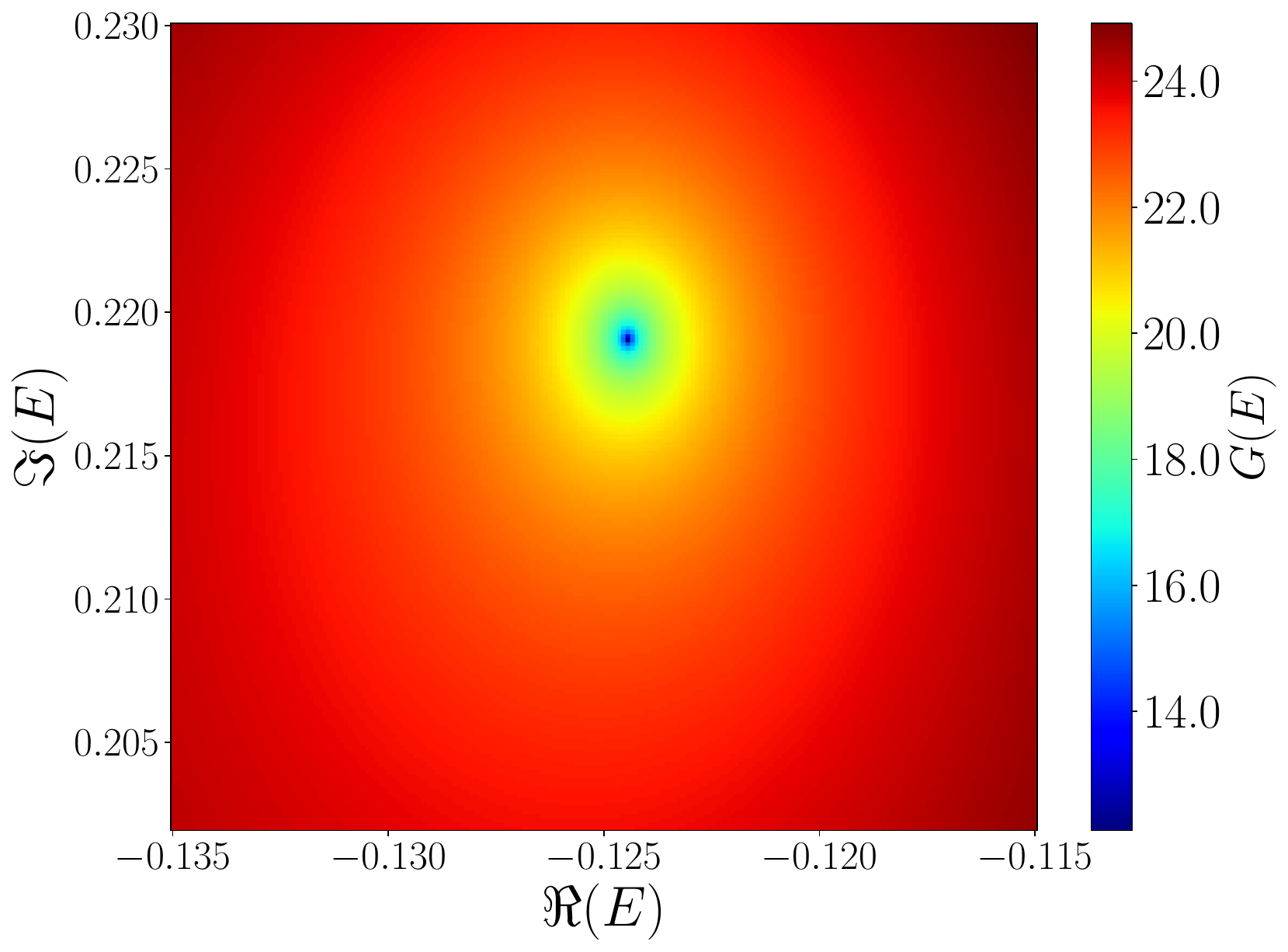}
  \caption{\label{fig:pinch-zoom}Real function $G(E)$ of Eq.~\eqref{eq:function} around the
    region of the breather energy. We observe clearly the appearance of a zero of the function at
    the value of the prescribed energy $E=1/8+\sqrt{3}\id/8$.}
\end{figure}
\medskip

\subsection{Breathers}
The breather is a time-oscillating and spatially localized soliton of the SG equation which can be described by an
angle $0<\mu<\pi/2$ in the complex plane of $\lambda$ and is associated with two complex values of
$\lambda_{1,2}=\sqrt{E}=\pm\frac{1}{2}\ed^{\pm \id\mu}$~\cite{Kivshar1991}. Its shape is given by
\begin{equation}\label{eq:kink2}
  \phi(x,t)=4\arctan\Bigg[\tan\mu\frac{\sin(t\cos\mu+\phi_0)}{\cosh(x\sin\mu)}\Bigg].
\end{equation}
Our calculation of the trace must be extended to complex values of the energy and we
shall be looking for the points where $\Im(\Tr \mathbf{M})=0$ and still
$\Re[\Tr(\mathbf{M})/2]=1$. Nevertheless, we find that the presence of breathers is easily identified by
simple inspection of the imaginary part of the trace, from which we numerically find and confirm
their energies.

We take the value of $\mu=\pi/3$ and apply the algorithm to a single breather. The search for the
eigenvalue is now extended into the complex plane by calculating the trace on a mesh of energies.
The real part of the trace of $\mathbf{M}$ is shown in Fig.~\ref{fig:trace-breather}(a), alongside the imaginary part
in Fig.~\ref{fig:trace-breather}(b). We observe an entangled pattern of values, which oscillate from
positive to negative {values}. The focus should be put on the imaginary part of the trace, which gives a
qualitative hint of the presence of a breather.

The breather changes the spectrum by ``pinching'' together two regions of negative (and positive)
values of the imaginary part of the trace. Without the presence of the soliton, the regions are well
separated and smooth (see Appendix~\ref{appendix}). The accurate location of the ``pinch" in the $E$-space
does not correspond exactly to its eigenvalue, and depends on the resolution of the mesh, but it serves
as a good qualitative indicator.

In order to locate the exact position of the breather eigenvalue we use the pinch as a first
guess, and focus on the function
\revise{\begin{equation}\label{eq:function}
  G=\big[\left|\Re(\Tr\mathbf{M})\right|-2\big]^2+\big[\Im(\Tr\mathbf{M})\big]^2
\end{equation}}
which will be identically zero when the real part is equal to $2$ and the imaginary part is zero. We
plot this function $G(E)$ in a zoomed-in region around the pinch in Fig.~\ref{fig:pinch-zoom}. As we
can see, there is a single minimum, which corresponds to the breather eigenvalue. By applying a
root-finding algorithm, we find the location of the root to be $E=-0.1268+0.2116\id$, which up to
our precision, is close to the expected value of $E=-\frac{1}{8}+\frac{\sqrt{3}}{8}\id$. For the
root-finding procedure, the position of the minimum found in Fig.~\ref{fig:pinch-zoom} was taken as
the initial guess.

\reviseCTP{In order to better quantify the detection of breathers instead of just relying on visual inspection, we  use Cauchy's argument principle which states that
  \begin{align}
    I=\frac{1}{2\id\pi}\oint_\gamma\frac{f'(z)}{f(z)}\,\mathrm{d} z,
  \end{align}
  for a holomorphic function $f(z)$ and a closed contour $\gamma$, is an integer equal to $Z$, the number of zeros of $f$ enclosed by $\gamma$ and counted with their multiplicities.  Here we compute the value of $I$ with $f(z)=\Tr\mathbf{M}(z)-2$. For the sake of
  simplicity and efficiency, we use a regular Cartesian grid of point of the complex $E$-plane,  a rectangle contour $\gamma$ based on the grid points and centered differences for the calculation of the complex derivatives along $\gamma$. We use trapezoidal rule in order to compute the integral. The value of the index is found to be $I=0.9995\approx 1$ (the numerical error stemming from the discretization). It corresponds to one breather, since, much like the kink, the order of the zero of $\Tr\mathbf{M}-2$ is 
  $1$. This method has also been tested in the case of multiple breathers and  found to be reliable: for instance, four breathers lead to $I=4$, provided $\gamma$ encloses the four zeros of $f$ (data not shown).}



\begin{figure*}[h!]
  \centering
  \quad(a) \hspace*{7.5cm}(b)
  
  \includegraphics[width=\columnwidth]{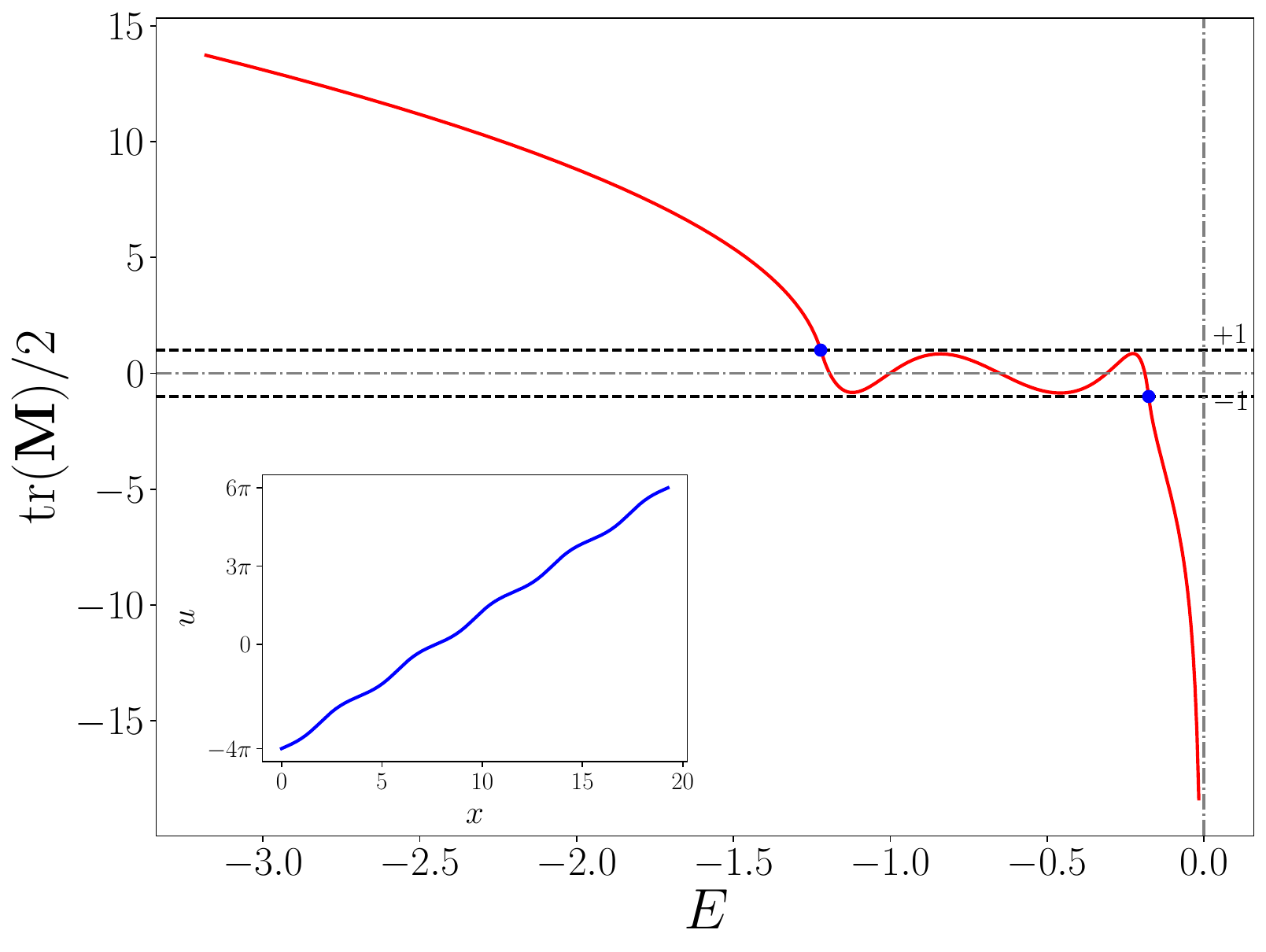}\;\;\includegraphics[width=\columnwidth]{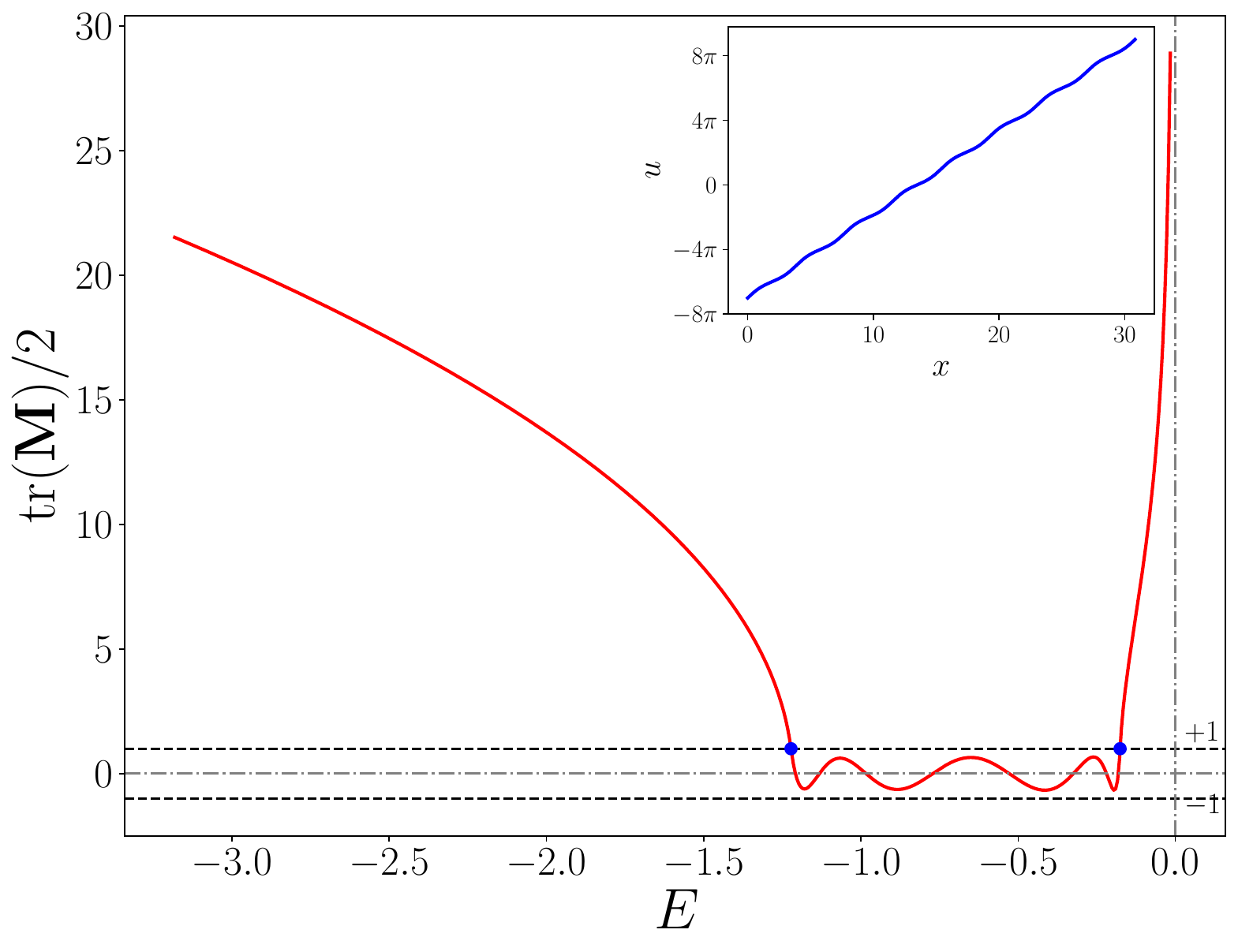}

  \caption{\label{fig:trace-kink-train}{Half-trace of the monodromy matrix $\mathbf{M}$ for a
    train of kinks (as in the initial condition in the inset), defined by two different energies. In
    addition to the crossings at the expected energies (blue dots), we observe that the trace is
    indeed zero between these two values. (a) Case of a train of $N=5$ kinks, (b) Case of a train of $N=8$ kinks. Depending on the parity of $N$, the value of the half-trace of the monodromy matrix $\mathbf{M}$ only crosses the value $+1$ of both the values $+1$ and $-1$.}}
\end{figure*}

\section{Periodic solutions: trains of kinks and radiative modes}\label{periodic_sol}

We now turn to the periodic solutions of the SG equation. We will be interested only in stable
solutions~\cite{Scott1969}. We will first test the method for nonlinear periodic solutions (cnoidal-like waves as a \reviseCTP{periodic} train of kinks) and then linear periodic solutions as radiative modes.

\subsection{\reviseCTP{Periodic} train of kinks}{We begin with the monotonically increasing kink train, which
is characterized by two energies $E_1<E_2<0$, and requires that $(E_1E_2)^{1/2}>1$. Its form is
given by Eq.~\eqref{formule_am}, and {the number of kinks that are contained in the domain sets the size of the latter } through Eq.~\eqref{eq:LmvN}. The velocity $v$ and elliptic parameter $m$ are given by
Eq.~\eqref{eq:train_velocity}. If we let $E_1$ and $E_2$ approach each other, the distance between
the kinks increases and we recover the {single} kink solution of Eq.~\eqref{eq:formule_kink}.}

{In order to test our direct scattering method on this solution, let us choose the kink train velocity to be $v=0.3$ and the elliptic parameter $m=0.8$. The half-trace $\mathbf{M}(E)$ is shown in
Fig.~\ref{fig:trace-kink-train} for two cases of a $N$-soliton train. We clearly observe that the half-trace is equal to $1$ at a low 
energy value $E_1$, then is equal to $-1$ at a higher energy value $E_2$ (Fig. \ref{fig:trace-kink-train}(a)). Conversely, when $N$ is even, the half-trace is equal to $+1$ at two
energy values (Fig. \ref{fig:trace-kink-train}(b)). In both cases, the numerically obtained values of the energies are found to be $E_1=-1.2222$ and
$E_2=-0.1757$, giving $v=0.2991$ and $m=0.797$. Moreover, and more interestingly, we observe that
the two eigenvalues are both located at the intersection with $+1$ for an even number of kinks, and that the value of the
half-trace remains within $\pm 1$ between the two and oscillates $N-1$ times. This is what distinguishes the kink train from
the pure kink solution.  }

\subsection{Radiative modes}
\begin{figure}[t!]
  \centering
  \includegraphics[width=\columnwidth]{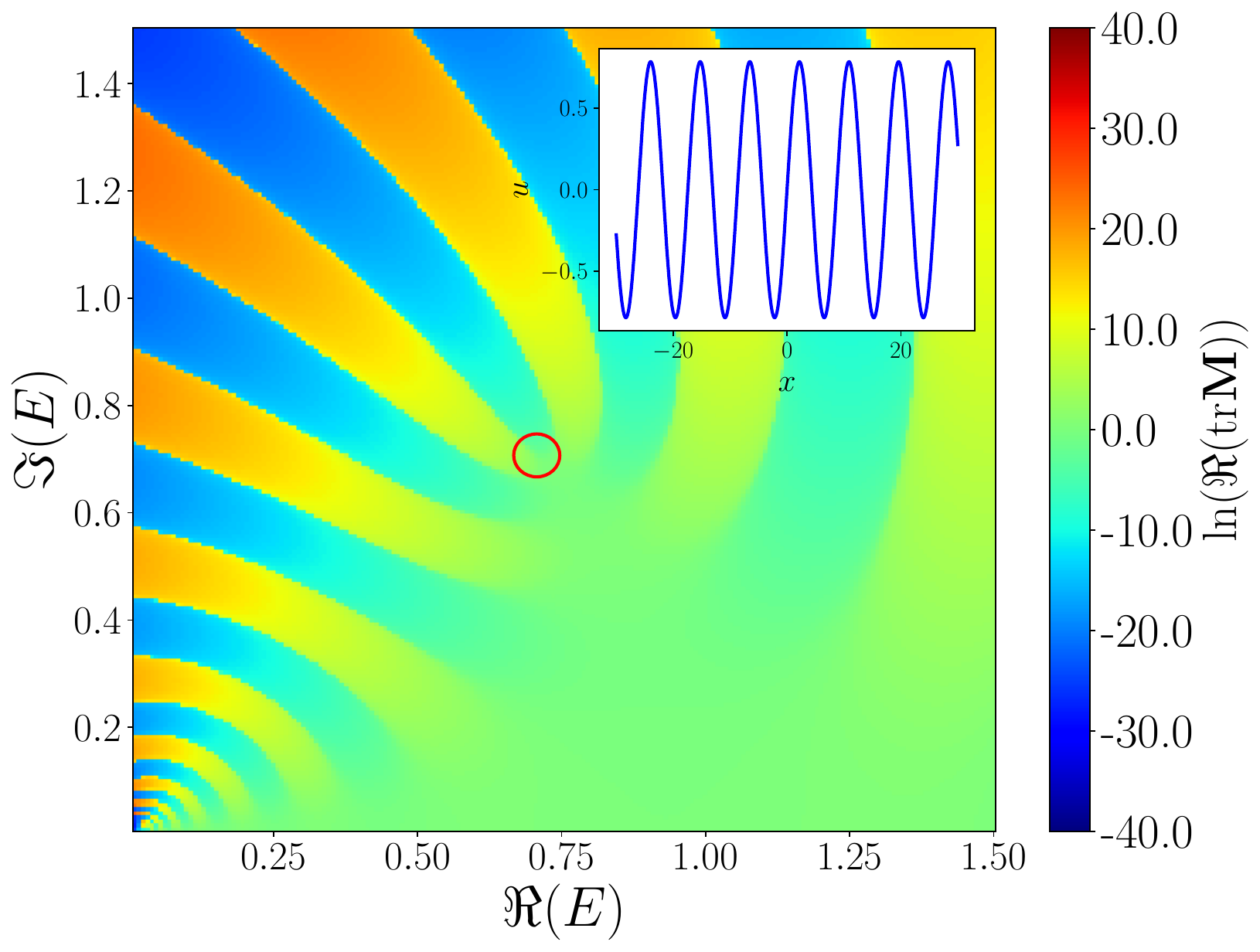}
  \caption{\label{fig:trace-wave}Real part of the trace of $\mathbf{M}$ for an oscillatory radiation excitation,
    as described by Eq.~\eqref{eq:wave}. The presence of such a mode distorts the trace at positive
    values of the real part of the energy, especially close to the prescribed mode energy. Inset: Waveform of Eq.~\eqref{eq:wave} used at $t=0$.}
\end{figure}

\begin{figure}[t!]
  \centering
  \includegraphics[width=\columnwidth]{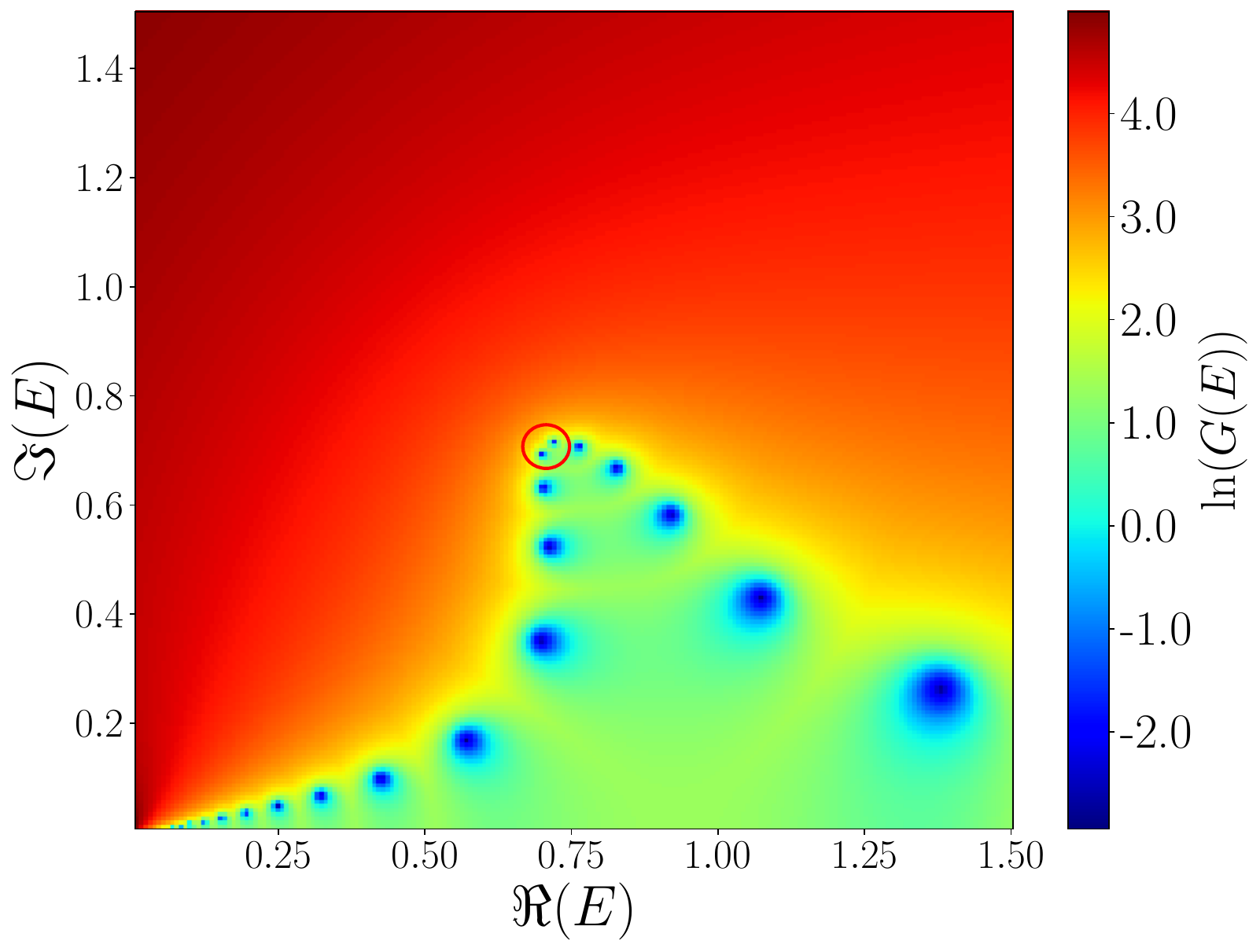}
  \caption{\label{fig:gfunc}Value of the function $G(E)$ of Eq.~\eqref{eq:function}. A zero of this function is
    found at the prescribed energy, as expected. We notice also additional zeros,
    corresponding to other radiative modes. Logscale colorbar.}
\end{figure}

Let us now turn to small-amplitude periodic solutions,
which will illustrate the dispersive or radiative components of the spectrum. This solution is
governed by two values of the energy $E_1=E_2^*=|E_1|\ed^{\id\alpha}$ located near the positive real
axis~\cite{Forest1982,Scott1969}. The form of the waves is given as
\begin{equation}\label{eq:wave}
  u=2\arcsin\left[\sqrt{m}\sn\left(\frac{x-vt}{\sqrt{1-v^2}},m\right)\right],
\end{equation}
where $\sn(x,m)$ is the Jacobi elliptic function of the first kind with elliptic parameter $m$.  The
velocity $v$ and parameter $m$ are slightly modified with respect to that of
Eq.~\eqref{eq:train_velocity} and read
\begin{gather}
  m=\sqrt{\frac{1-\cos{\alpha}}{2}},\\ v=\frac{1+4(E_1E_2)^{1/2}}{1-4(E_1E_2)^{1/2}}.
\end{gather}
The amplitude of the wave is controlled through the angle of the energy $\alpha$. We select a value
of the energy with $\alpha=\pi/4$ and $|E_1|=1$ and focus on the part of the complex plane with
positive real values of $\Tr(\mathbf{M})$. The inset of Fig.~\ref{fig:trace-wave} shows the
  waveform of Eq.~\eqref{eq:wave} at $t=0$. {In Fig.~\ref{fig:trace-wave}, we only show }  the real
part of the trace of the matrix $\mathbf{M}$ since there is no significant qualitative difference
with its imaginary part. Similarly to the breather of Sec.~\ref{sol}, a distortion of the trace in
the complex plane is observed, and a similar pinching is found close to the energy of the traveling
kink-train, in particular when compared to the trace of $\mathbf{M}$ corresponding to a
low-amplitude sine wave (see Appendix~\ref{appendix}). In order to evidence the presence of
the prescribed energies {in a clearer way}, we once more make use of the function $G(E)$ of Eq.~\eqref{eq:function},
which is plotted in Fig.~\ref{fig:gfunc}. We clearly see that there is a zero at the position of the
prescribed energy, circled in red, as predicted by Floquet theory. Other zeros, corresponding to
similar radiative modes, are also present.


\section{Conclusion}\label{conclusion}
We have proposed a numerical method for the direct scattering problem of the periodic SG
equation, {based on }a technique previously used in the case of the KdV equation. Applied to the
SG equation, this method leads to obtaining the trace of the scattering matrix and finding
the eigenvalues associated with the nontrivial direct scattering problem.  We then computed the
scattering matrix for several basic soliton solutions of the SG equation, for which the
associated eigenvalues are either real (topological solitons as kinks, anti-kinks, and kink trains)
or complex in the energy complex plane {in the case of} non-topological solitons solutions (breathers) and
periodic solutions.

This method avoids {the direct and tedious solving of } the
nonlinear eigenvalue problem and can be easily applied to a given time
series. This could potentially provide {meaningful
  informations in} experiments, as in the cases of both the NLS and
KdV equations. Furthermore, the dynamics of a large number of
solitons, such as soliton gas, is a field of growing interest in the
study of integrable turbulence~\cite{Suret2023}, and the SG equation
could provide richer dynamics with its topological soliton solutions.

\bmhead{Acknowledgments}
This work is supported by the French National Research Agency (ANR SOGOOD project
No. ANR-21-CE30-0061-04), and by the Simons Foundation MPS No 651463-Wave Turbulence.

\bmhead{Data Availability Statement}
No Data associated in the manuscript.
  
  
\begin{appendices}
\section{}\label{appendix}

\begin{figure*}[t!]
  \centering
  (a) \hspace*{7cm}(b)
  
\includegraphics[width=\columnwidth]{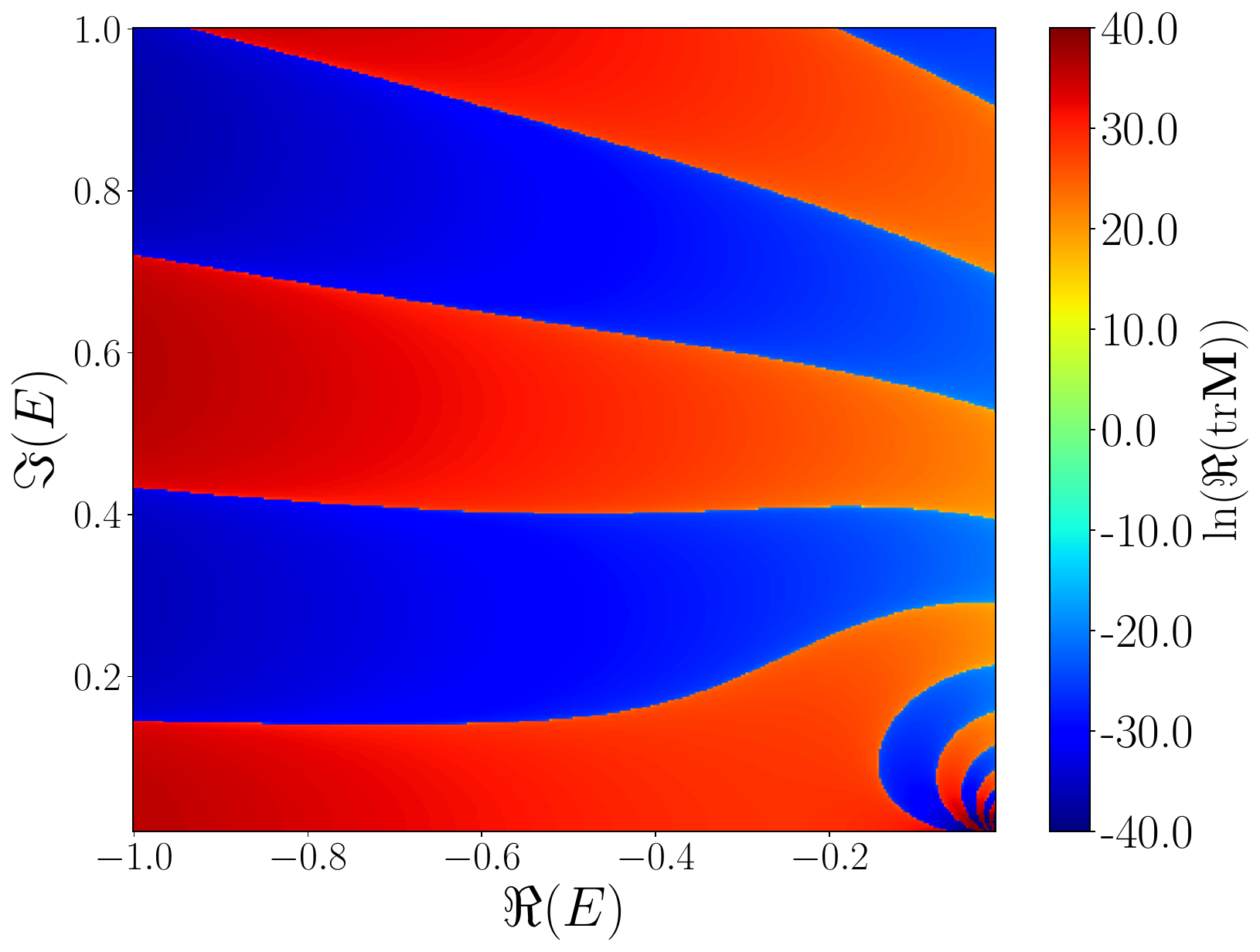}
\includegraphics[width=\columnwidth]{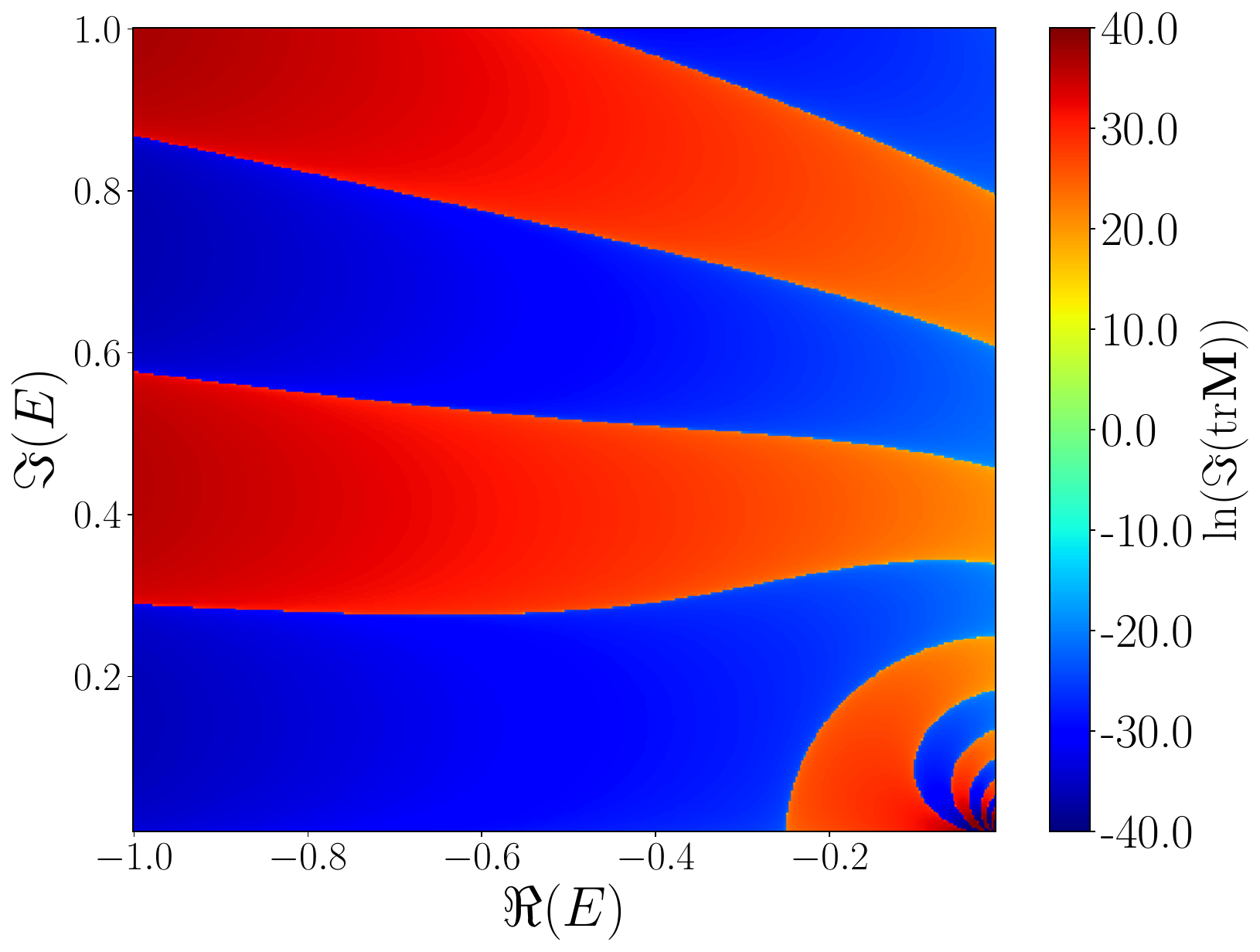}
  \caption{\label{fig:sine-neg}(a) Real part of the trace of $\mathbf{M}(E)$ in the
    complex plane of energy, for negative $\Re(E)$, for a low-amplitude sine wave. (b) Imaginary  part of the trace of
    $\mathbf{M}(E)$ for the same wave.}
\end{figure*}

\begin{figure*}[t!]
  \centering
  (a) \hspace*{7cm}(b)
  
  \includegraphics[width=\columnwidth]{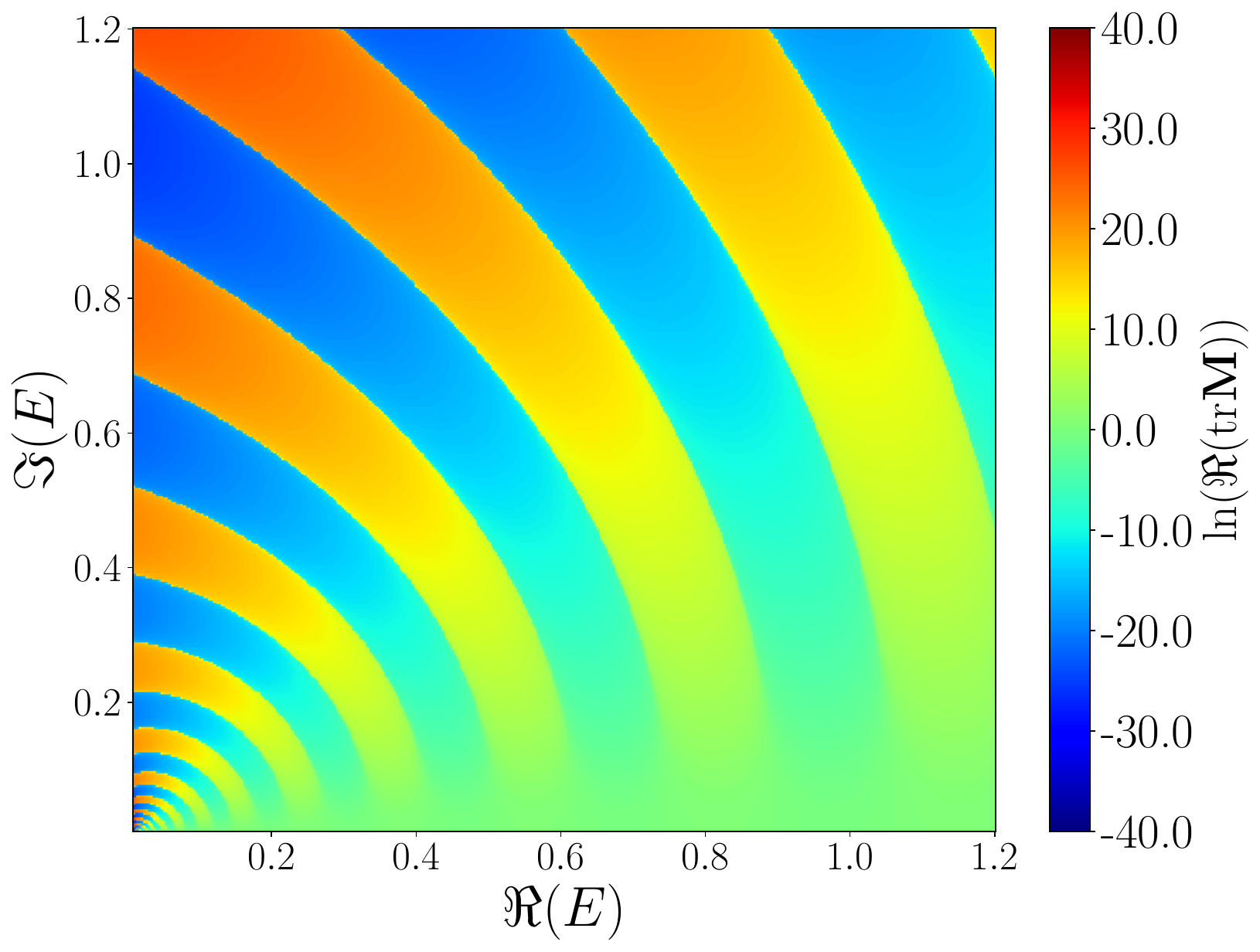}
  \includegraphics[width=\columnwidth]{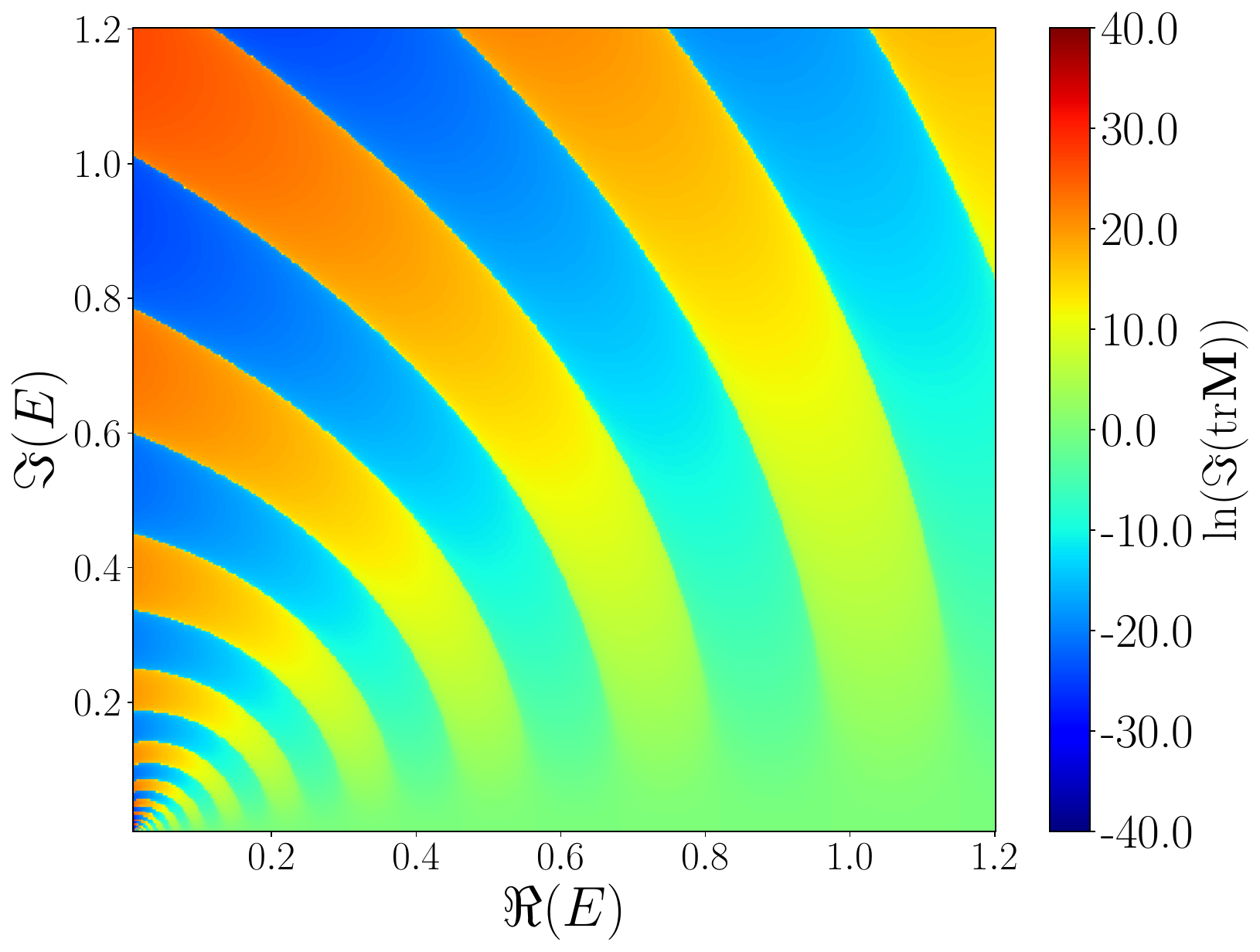}
  \caption{\label{fig:sine-pos}%
    (a) Real part of the trace
    of $\mathbf{M}(E)$    in the complex
    plane of energy, for positive $\Re(E)$, for a low-amplitude sine wave.
    (b) Imaginary part of the trace of $\mathbf{M}(E)$  for the same wave. }
\end{figure*}

As a comparison with the results on the different solitons and
periodic solutions to the pSG equation discussed in the main text, we here
include the half-trace obtained for a sine wave of amplitude
$A=0.01$, with $k=2$ and $\omega=1$ on a domain of length $L=60$. The
real and imaginary parts of the trace are shown in
Fig.~\ref{fig:sine-neg} for the negative real part of the energy plane,
where we observe that there is no pinching. In Fig.~\ref{fig:sine-pos}
we show the same quantity for the positive real part of the energy plane.

\textcolor{red}{\section{}}\label{appendix2}

\begin{figure*}[t!]
  \centering
  \qquad(a) \hspace*{7.5cm}(b)
  
  \includegraphics[height=.72\columnwidth]{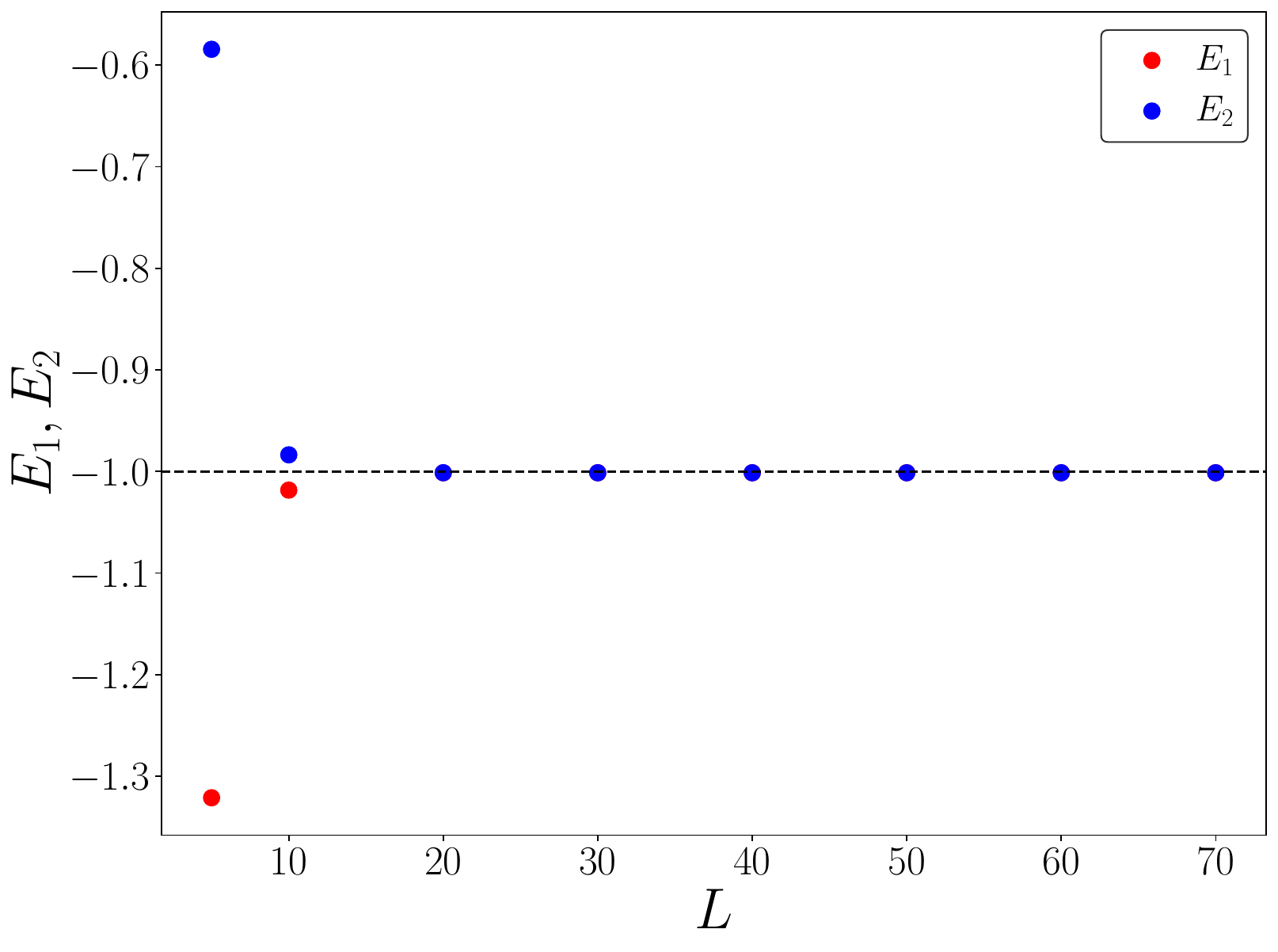}
  \quad
  \includegraphics[height=.72\columnwidth]{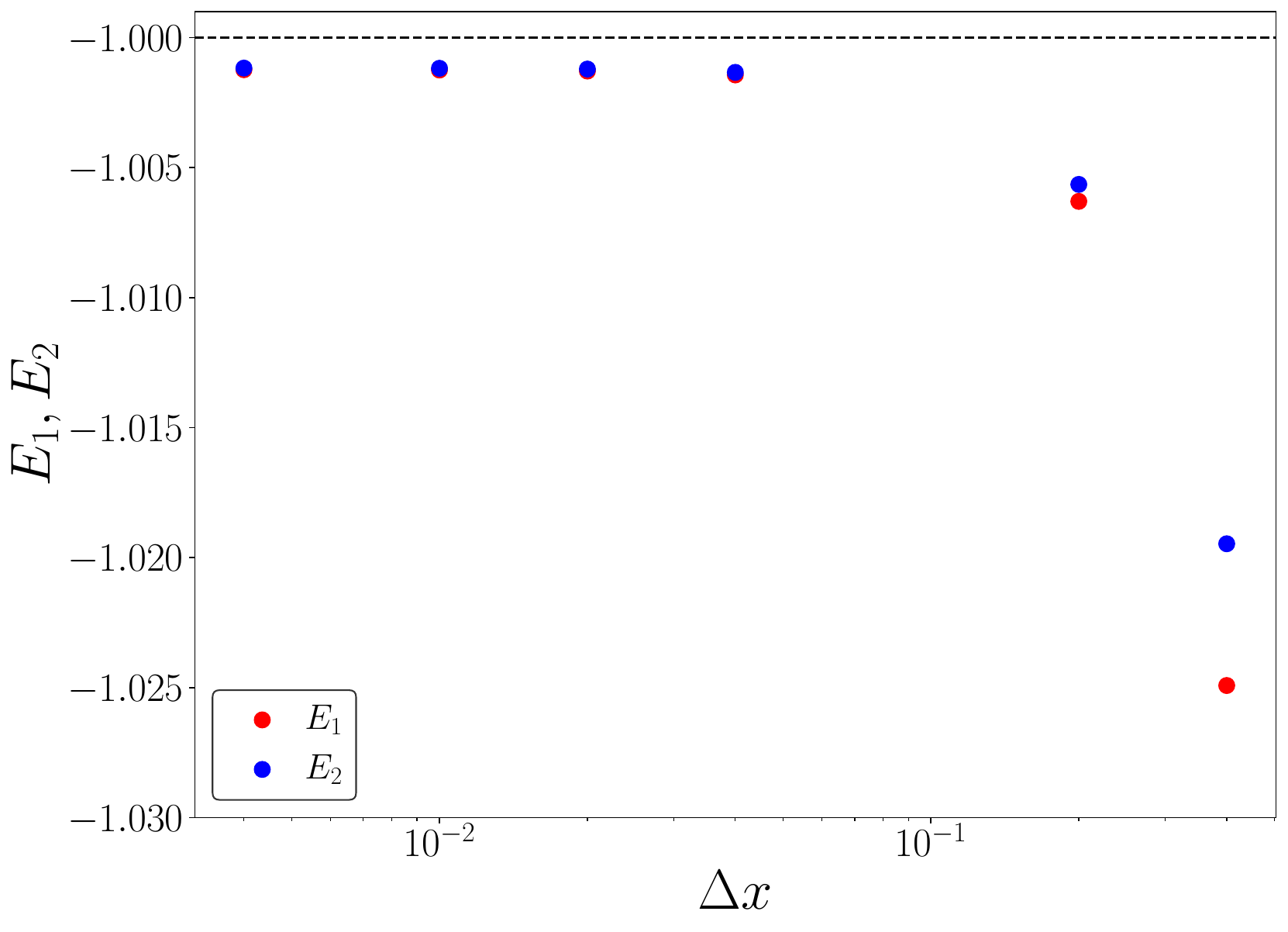}
\caption{\label{fig:error}(a) The eigenvalues of the truncated kink with energy $E=1$ for
  different values of the domain size. (b) Eigenvalues of the truncated kink on a domain with fixed $L=20$
  for different values of the discretization $\Delta x$.}
\end{figure*}

\revise{In this appendix, we discuss how the domain size $L$ and the spatial discretization $\Delta x$ affect the eigenvalues of a
single infinite-line truncated kink soliton. We start varying $L$ while keeping $\Delta x$ fixed at $0.02$. Its energy has been
chosen to be $E=1$. We can see in Fig.~\ref{fig:error}(a) that as the domain size $L$ is increased, the
results become more accurate. This is due to the fact that the errors due to periodicity (i.e. the
mismatch at the domain ends) decreases.  In Fig.~\ref{fig:error}(b), we have kept the domain size value at
$L=20$ and changed the discretization $\Delta x$. As expected, by decreasing the value of
$\Delta x$, the energies converge to a single value. }

\end{appendices}

\medskip



\end{document}